\input harvmac.tex
\input epsf
\noblackbox
\newcount\figno
 \figno=0
 \def\fig#1#2#3{
\par\begingroup\parindent=0pt\leftskip=1cm\rightskip=1cm\parindent=0pt
 \baselineskip=11pt
 \global\advance\figno by 1
 \midinsert
 \epsfxsize=#3
 \centerline{\epsfbox{#2}}
 \vskip 12pt
 {\bf Fig.\ \the\figno: } #1\par
 \endinsert\endgroup\par
 }
 \def\figlabel#1{\xdef#1{\the\figno}}
\newdimen\tableauside\tableauside=1.0ex
\newdimen\tableaurule\tableaurule=0.4pt
\newdimen\tableaustep
\def\phantomhrule#1{\hbox{\vbox to0pt{\hrule height\tableaurule width#1\vss}}}
\def\phantomvrule#1{\vbox{\hbox to0pt{\vrule width\tableaurule height#1\hss}}}
\def\sqr{\vbox{%
  \phantomhrule\tableaustep
  \hbox{\phantomvrule\tableaustep\kern\tableaustep\phantomvrule\tableaustep}%
  \hbox{\vbox{\phantomhrule\tableauside}\kern-\tableaurule}}}
\def\squares#1{\hbox{\count0=#1\noindent\loop\sqr
  \advance\count0 by-1 \ifnum\count0>0\repeat}}
\def\tableau#1{\vcenter{\offinterlineskip
  \tableaustep=\tableauside\advance\tableaustep by-\tableaurule
  \kern\normallineskip\hbox
    {\kern\normallineskip\vbox
      {\gettableau#1 0 }%
     \kern\normallineskip\kern\tableaurule}%
  \kern\normallineskip\kern\tableaurule}}
\def\gettableau#1 {\ifnum#1=0\let\next=\null\else
  \squares{#1}\let\next=\gettableau\fi\next}
\tableauside=1.0ex \tableaurule=0.4pt

\def\frac#1#2{{#1\over #2}}

\lref\atishone{
  A.~Dabholkar,
  ``Exact counting of black hole microstates,''
  {\tt hep-th/0409148}.
}

\lref\atishtwo{
  A.~Dabholkar, F.~Denef, G.~W.~Moore and B.~Pioline,
  ``Exact and asymptotic degeneracies of small black holes,''
  {\tt hep-th/0502157}.
}

\lref\aosv{
M.~Aganagic, H.~Ooguri, N.~Saulina and C.~Vafa,
``Black holes, q-deformed 2d Yang-Mills, and non-perturbative topological
strings,''
{\tt hep-th/0411280}.
}

\lref\vafaqcd{
C.~Vafa,
``Two-dimensional Yang-Mills, black holes and topological strings,''
{\tt hep-th/0406058}.
}

\lref\hh{J.~B.~Hartle and S.~W.~Hawking,
``Wave function of the universe,''
Phys.\ Rev.\ D {\bf 28}  (1983)
 2960.
}

\lref\wignerfun{E.~P.~Wigner,  Phys.\ Rev.\
``On the quantum correction for thermodynamic equilibrium,''
{\bf 40} (1932) 749.}

\lref\husimifun{K.~Husimi,
``Some formal properties of the density matrix,''
 Proc.\ Phys.\
Math.\ Soc.\ Japan, {\bf 22} (1940) 264.}

\lref\FGK{
S.~Ferrara, G.~W.~Gibbons and R.~Kallosh,
``Black holes and critical points in moduli space,''
Nucl.\ Phys.\ B {\bf 500} (1997) 75;
{\tt hep-th/9702103}.
}

\lref\Liu{
H.~Liu, G.~W.~Moore and N.~Seiberg,
``Strings in a time-dependent orbifold,''
JHEP {\bf 0206} (2002)  045;
{\tt hep-th/0204168}.
}
\lref\denef{ F.~Denef,
``Supergravity flows and D-brane stability,''
JHEP {\bf 0008}  (2000)  050;
{\tt hep-th/0005049}.}
\lref\osv{H.~Ooguri, A.~Strominger, and C.~Vafa,
``Black hole attractors and the topological string,''
Phys.\ Rev.\ D70 (2004) 106007, {\tt hep-th/0405146}.
} \lref\V{C.~Vafa, ``Two-dimensional Yang-Mills, black holes and
topological strings,'' {\tt hep-th/0406058}.} \lref\bcov{ M.
Bershadsky, S. Ceccoti, H. Ooguri and C. Vafa, ``Kodaira-Spencer
theory of gravity and exact results for quantum string
amplitudes,'' Commun. Math. Phys. 165 (1994) 311, {\tt
hep-th/9309140}.} \lref\naret{ I.~Antoniadis, E.~Gava,
K.~S.~Narain and T.~R.~Taylor, ``Topological amplitudes in string
theory,'' Nucl.\ Phys.\ B413 (1994) 162, {\tt hep-th/9307158}.
}
\lref\ovc{H.~Ooguri and C.~Vafa, ``Two-dimensional black hole
and singularities of CY manifolds,'' Nucl.Phys. B463 (1996) 55-72,
{\tt hep-th/9511164}.
} \lref\INOV{A.~Iqbal, N.~Nekrasov, A.~Okounkov, and C.~Vafa,
``Quantum foam and topological strings", {\tt hep-th/0312022}.}
\lref\OPN{D.~Maulik, N.~Nekrasov, A.~Okounkov, and
R.~Pandharipande, ``Gromov-Witten theory and Donaldson-Thomas
theory," math.AG/0312059 } \lref\NV{A.~Neitzke and C.~Vafa,
     ``${\cal N}=2$ strings and the twistorial Calabi-Yau," {\tt hep-th/0402128}.}
\lref\NOV{N. Nekrasov, H. Ooguri, C. Vafa, ``S-duality and topological
strings," JHEP\ 0410 (2004) 009, {\tt hep-th/0403167}.
}
\lref\OOV{H. Ooguri and C. Vafa, ``Knot invariants and topological
strings," Nucl.\ Phys.\ B577 (2000) 419-438, {\tt hep-th/991213}.
}
\lref\dewit{G.~Lopes Cardoso, B.~de~Wit and T.~Mohaupt,
``Macroscopic entropy formulae and non-holomorphic corrections for
supersymmetric black holes,'' Nucl.\ Phys.\ B {\bf 567} (2000) 87;
{\tt hep-th/9906094}.}
\lref\fks{S.~Ferrara, R.~Kallosh and A.~Strominger,
``${\cal N}=2$ extremal black holes'', Phys. Rev. D52
512 (1995), {\tt hep-th/9508072}. }
\lref\asm{A.~Strominger, ``Macroscopic Entropy of ${\cal N}=2$ black holes,''
 Phys. Lett. B383 (1996) 39,  {\tt hep-th/9602111}. }
\lref\ferrara{S.~Ferrara and R.~Kallosh, ``Supersymmetry and
Attractors,'' Phys.\ Rev.\ D {\bf 54}, 1514 (1996);
{\tt hep-th/9602136}.} \lref\strominger{A.~Strominger,
``Macroscopic Entropy of $N=2$ Extremal Black Holes,'' Phys.\
Lett.\ B {\bf 383}, 39 (1996) [arXiv:hep-th/9602111].}
\lref\Witten{E.~Witten, ``Quantum background independence in
string theory,'' {\tt hep-th/9306122}.} \lref\goldstein{H.~Goldstein, 
C.~P.~Poole, and J.~L.~Safko, "Classical Mechanics," 
(Addison-Wesley, 2002).}
\lref\DVV{R.~Dijkgraaf, E.~Verlinde and M.~Vonk,``On the partition
sum of the NS five-brane,'' {\tt hep-th/0205281}.}
\lref\EV{E.~Verlinde, ``Attractors and the holomorphic anomaly,''
{\tt hep-th/0412139}.} \lref\moore {G. Moore "Arithmetic and Attractors
", [archive:hep-th/9807087]}
\lref\topm{R. Dijkgraaf, S. Gukov, A. Neitzke, C.  Vafa,
``Topological M-theory as unification of form theories of gravity,"
{\tt hep-th/0411073}.}
\lref\atish{A. Dabholkar, "Exact Counting of Black Hole
Microstates," [archive: hep-th/0409148]}
\lref\dewitold{K.~Behrndt, G.~Lopes Cardoso, B.~de Wit, 
R.~Kallosh, D.~L\"ust and T.~Mohaupt,
"Classical and quantum ${\cal N}=2$ supersymmetric black holes", 
Nucl. Phys. {\bf B488}, 236 (1997); {\tt hep-th/9610105}}
 \lref\tye{H. Firouzjahi, S. Sarangi, S.-H. Tye,
"Spontaneous Creation of Inflationary Universes 
and the Cosmic Landscape'', JHEP {\bf 0409}, 060 (2004); {\tt hep-th/0406107}}
\lref\Koba{
  A.~Kobakhidze and L.~Mersini-Houghton,
  ``Birth of the universe from the landscape of string theory,''
  {\tt hep-th/0410213}.
}

\Title{
  \vbox{\baselineskip12pt \hbox{hep-th/0502211}
  \hbox{CALT-68-2543}
\hbox{HUTP-05/A005}
\hbox{ITFA-2005-05}
  \vskip-.5in}
}{\vbox{
  \centerline{Hartle-Hawking   Wave-Function for Flux Compactifications:}
\vskip .2in \centerline{The Entropic Principle} }}
\centerline{Hirosi Ooguri,$^1$~ Cumrun Vafa,$^2$~ and  Erik
Verlinde$^3$}
\bigskip\bigskip\smallskip
\centerline{$^1$ \it California Institute of Technology
452-48, Pasadena,
CA 91125, USA}
\vskip .05in
\centerline{$^2$ \it Jefferson Physical Laboratory,
Harvard University, Cambridge, MA 02138, USA}
\vskip .05in
\centerline{$^3$ \it Institute for Theoretical
Physics, University of Amsterdam,}
\centerline{\it Valckenierstraat 65, 1018 XE Amsterdam, The Netherlands}
\medskip
\medskip
\medskip
\vskip .5in
We argue that the topological string partition function, which has been known
to correspond to a wave-function, can be interpreted as an exact ``wave-function
of the universe'' in the mini-superspace sector of
physical superstring theory.  This realizes the idea
of Hartle and Hawking in the context of string theory, including
all loop quantum corrections.  The mini-superspace approximation is
justified as an exact description of BPS quantities.  Moreover this proposal
leads to a conceptual explanation of the recent
observation that the black hole entropy is the square
of the topological string wave-function.  This wave-function can be
interpreted in the context of flux compactification of all spatial
dimensions as providing a physical probability distribution
on the moduli space of string compactification.
Euclidean time is realized holographically in this
setup.
\Date{February, 2005}
\newsec{Introduction}

In string compactifications we typically start
by assuming that the universe includes a non-compact macroscopic
spacetime.  Even though this is a natural assumption to make
in the present epoch of cosmology, it is far less clear whether it
is a good assumption in the very early universe.  In particular one
can imagine a universe arising from a microscopic/Planckian size
compact space.  In fact this seems like a more reasonable assumption in the context
 of cosmology.

On the other hand if we consider complete compactification of all spatial
directions in string theory, we should not be able to freeze the moduli
of string theory.  More precisely we would expect to have a wave-function
on the moduli space of string compactifications, whose amplitudes
is peaked at the values
of moduli where the classical equations of motion would be obeyed.
Such a notion of wave-function seems necessary if one is to have a physical
framework to compare the relative probabilities of different
string compactifications.

Appealing as this idea is, there has been almost no progress
in this direction in string theory.
 Our main aim in this paper is to present a modest
progress in this direction for compactifications which preserve
supersymmetry, where exact computations are possible.  
As a by-product, we are also
able to make contact with the recent result that the exact black hole
entropy is given by the norm square of the topological string
partition function \osv. This connection arises since we are able to
identify the partition function
as the wave-function of the universe in the mini-superspace description,
analogous to the Hartle-Hawking construction \hh.
The mini-superspace approximation is
justified as an exact description of BPS quantities.
Moreover we find
the precise form of this wave-function which includes corrections
to all orders in the string perturbation theory.

We believe this work is a first step toward developing
a more realistic quantum cosmology within string theory. Given
the recent exciting observational data, such a development
would be rather timely.

The original Hartle-Hawking no-boundary proposal gives a
construction of a special solution to the Wheeler-De~Witt (WDW)
equation to the Einstein gravity with a positive cosmological
constant on a three-sphere $S^3$ in terms of a formal Euclidean
functional integral over a four-dimensional ball $B_4$, which is
bounded by the $S^3$. There are other four-dimensional spaces
bounded by $S^3$, and they would give different solutions to the
WDW equation. Hartle and Hawking proposed that the particular
solution associated to $B_4$ plays a distinguished role as a wave
function of the universe. This wave-function was used to set the
initial boundary condition of the universe, which would expand
exponentially due to the positive cosmological constant. Since the
Euclidean functional integral in the Einstein gravity is not
well-defined by itself, Hartle and Hawking considered a simplified
system where the degrees of freedom are reduced to zero modes on
$S^3$ -- this is called the mini-superspace approximation. Their
construction of the wave-function can be made more rigorous if the
computation can be formulated in the string theory, where a full
quantum theory of gravity is available.

In this paper, we will find a string theoretical realization of a
similar construction for certain flux compactifications, where the
cosmological constant is negative\foot{The 
Hartle-Hawking wave function for flux compactifications in de Sitter space was
discussed in \refs{\tye,\Koba}.}   
Motivated by the recent result
on the entropy of Calabi-Yau black holes \osv , we are
particularly interested in compactification of type II string on
$S^2$ times a Calabi-Yau three-fold $M$ with fluxes so that we
have a $(1+1)$-dimensional spacetime with a negative cosmological
constant. To relate it to the black hole computation, we will
further compactify the remaining one spatial dimension to a circle
$S^1$ with the periodic (supersymmetry preserving) boundary
condition on the fermions. To apply the Hartle-Hawking
prescription, we look for a space that is bounded by $S^1 \times
S^2 \times M$ and that supports a classical solution to the string
equations of motion with an Euclidean signature metric. In the
following, we will construct such a space and show that the
topological string partition function gives the Hartle-Hawking
type wave-function associated to this geometry.  The black hole
microstates are holographically dual to this geometry by exchange
of a spatial and temporal direction in Euclidean space
(target space analog of open/closed channel duality on the worldsheet).  
In particular the Euclidean gauge theory describing
the black hole microstates leads to a holographic realization of Euclidean
time in this setup.
We also discuss
some possible generalizations to other compactifications of string theory.

The organization of this paper is as follows:  In section 2
we recast the black hole attractor mechanism as the extremization
of flux generated superpotential upon compactifications on $S^2\times M$.  In
section 3 we introduce the notion of the mini-superspace for the problem at hand.
In section 4 we review attractor flows in the context of black hole
fluxes.  In section 5 we setup the formulation of WDW equation
in the BPS subsector of the mini-superspace.  Here
we propose a semi-classical notion of the wave function.  In section 6
we make our proposal for the full wave function, where we make
contact with the result of \osv , in relating the
black hole entropy to topological string wave-function.
%
In section 7 we
briefly
consider more general compactifications.
We conclude this paper with a discussion of some directions
for future research in section 8.

\newsec{Flux Compactifications and the Attractor Mechanism}

In this section we discuss attractor mechanism for
black holes \refs{\fks,\asm}
in a way to make contact
with some of the recent discussions on compactifications of string
theory \ref\fluxc{S.~Kachru, R.~Kallosh, A.~Linde and S.~P.~Trivedi,
``De Sitter vacua in string theory,''
Phys.\ Rev.\ D {\bf 68}, 046005 (2003);
{\tt hep-th/0301240}.
}.  The most thoroughly studied
class involves compactifications
on Calabi-Yau manifolds with fluxes turned on, similar to what we will be
 considering in the context of black holes.  However, it is useful
to start not from the black hole perspective, but
from the perspective of compactification to $(1+1)$ dimensions.

\subsec{Compactification on Calabi-Yau times $S^2$}

Suppose we consider the compactification of type IIB superstrings to
two dimensions on a Calabi-Yau threefold $M$ times a 2-sphere $M\times S^2$.
Turn on 5-form fluxes for the RR 5-form field strength
to be
\eqn\fiveflux{ F_5= F_3\wedge \omega,}
where $\omega$ is a unit volume form on $S^2$, and $F_3$ is a 3-form
on $M$.  Choosing an integral basis of magnetic/electric $H^3(M)$ as
$\{ \alpha_I, \beta^J\}_{I=0,...,h^{2,1}}$, we write
\eqn\threeform{ F_3=\sum_I
\left( p^I \alpha_I +q_I \beta^I \right)
.}
It is possible, as in \ref\gvw{
S.~Gukov, C.~Vafa and E.~Witten,
``CFT's from Calabi-Yau four-folds,''
Nucl.\ Phys.\ B {\bf 584}, 69 (2000),
[Erratum-ibid.\ B {\bf 608}, 477 (2001)];
{\tt hep-th/9906070}.
},  to write down a superpotential
whose extremization
leads to the condition for $(2,2)$ supersymmetry in $d=2$:
\eqn\gvwpotential{W=\int_{M\times S^2} F_5\wedge \Omega ,}
where $\Omega$ is the holomorphic 3-form on the Calabi-Yau
three-fold.  To deduce
this superpotential, note that this
superpotential is consistent with the tension (BPS mass in 1 dimension)
of the domain wall D3 brane
which wraps a 3-cycle in the Calabi-Yau and changes the $F_5$ flux.

The condition for extremization of $W$ and preserving supersymmetry
is
\eqn\extre{DW=0.}
Geometrically it is the following variational problem:
the complex structure of the Calabi-Yau
are field variables and we can look for extrema of $W$ with respect
to their variation.
Denote variation of $\Omega$ in arbitrary direction of $H^{2,1}(X)$
by $\delta \Omega$. The supersymmetry condition \extre\ is equivalent to
$$\int_{M\times S^2} \delta \Omega \wedge F_5 =\int_{M} \delta \Omega \wedge
F_3=0$$
Since $F_3$ is also real, this implies that\foot{This extremization
condition is similar to the one encountered in a different context \ref\GKP{S.~B.~Giddings,
S.~Kachru and J.~Polchinski,
``Hierarchies from fluxes in string compactifications,''
Phys.\ Rev.\ D {\bf 66}, 106006 (2002);
{\tt hep-th/0105097}.
}.}
 $F_3\in H^{3,0}+H^{0,3}$.
Using
the fact that there is only one element in $H^{3,0}$ represented by $\Omega$,
and using the reality condition for $F_3$ we deduce that
$$F_3={\rm Re}\left( C\ \Omega \right), $$
for some complex number $C$.   In other words
\eqn\attraction{ p^I={\rm Re}\left( C X^I\right)
 , ~~ \ q_I={\rm Re}\left( C F_I\right), }
where
$$ X^I = \int_{A_I} \Omega, ~~F_I = \int_{B^I} \Omega, $$
and $(A_I, B^I)$ are 3-cycles on $M$ that are
dual to the 3-forms $(\alpha_I, \beta^I)$.
With the complex
structure of the Calabi-Yau satisfying \attraction ,
supersymmetry is preserved with a suitable
choice of metric in 2 spacetime dimensions, $i.e.$ the $AdS_2$ metric.
For such a complex structure,
the superpotential $W$ is not zero, but it is proportional to
$\int_M \Omega \wedge {\overline \Omega}$.
In addition, the size of $S^2$ is also
determined by the supersymmetry condition, and we find that
\eqn\spherearea{ {\rm Area}(S^2)= \pi C\bar{C}
 \int_M \Omega \wedge {\overline \Omega}.}
In the following, we will set the constant $C$ to be equal to $1$
with a suitable rescaling of $\Omega$.

This is exactly the content of the attractor mechanism. To case it
in the standard description of the black hole attractor, consider D3
branes wrapping $q_I$ times on $A_I$ and $p^I$ times
on $B^I$. This gives rise to
a supersymmetric black hole in four dimensions, whose BPS mass $M_{BPS}$
is given by
\eqn\bpsmass{ M_{BPS}^2 = K^{-1} |W|^2,}
where the exponentiated K\"ahler potential $K$ is given by
\eqn\whatk{ K = i(\bar X^I F_I - X^I \bar F_I)
= -2\ {\rm Im}\tau_{IJ} X^I \bar X^J, }
and $W$ is the superpotential \gvwpotential, which can also be expressed as
\eqn\bpsmass{
\eqalign{ W & =  q_I \int_{A_I} \Omega - p^I
\int_{B^I} \Omega \cr
& = q_I X^I - p^I F_I.}}
Here and in the following, we use the Einstein convention and always
sum over repeated indices. The period matrix $\tau_{IJ}$ is defined by
\eqn\whattau{ \tau_{IJ}= {\partial F_J\over \partial X^I}. }
Extremizing $M_{BPS}$ with respect to the complex moduli of $X$,
with the constraint that $K$ is constant, reproduces the attractor
equations \attraction , where the constant $C$ is identified as a
Lagrange multiplier for the constraint $K={\rm const}$.

For the purpose of this paper, it is useful to describe the attractor
mechanism using the action principle \dewitold. Variation of the
action %
\eqn\action{\eqalign{ S=& -{\pi \over 4}\left(K(X,\bar X) + 2i
W(X)- 2i\bar W(\bar X) \right) \cr =& -{\pi \over 4}\left( \int
\Omega \wedge {\bar \Omega}+\int (\Omega +\bar \Omega) \wedge F_3
\right),}}
in arbitrary directions of $H^3(M)$ reproduces the attractor
equations \attraction .
To see this, note that
variation of $\Omega$ in the $H^{2,1}$ direction, $i.e.$ $DW=0$,
does not change the value of $K$ because
$\delta K=\int \delta \Omega \wedge {\bar \Omega}=0$.
Thus, such a variation can be regarded as a variation of the
action $S$ where $K$ is fixed. On the other hand,
varying $\Omega$ parallel to itself ($i.e.$ in the $H^{3,0}$ direction)
precisely leads to \attraction\ in the gauge $C=1$.
Another way to see this is to take a direct variation of $S$ with
respect to $X$. The extremum of $S$ is given by
\eqn\extremum{ X^I = \left({-i \over {\rm Im}\tau}\right)^{IJ}
(q_J -\bar \tau_{JK} p^K).}
Note that this is still a non-linear equation since $\tau_{IJ}$
in the right-hand side is evaluated at $X$ as in \whattau .
Taking the real and imaginary parts of this equation, one
reproduces the attractor equation \attraction\ with $C=1$.
For later convenience, we define $X_{p,q}^I$ by
\eqn\xps{ X_{p,q}^I = \left({-i\over {\rm Im}\tau}\right)^{IJ}
(q_J - \bar\tau_{JK} p^K)_{\bigl|{\rm attractor}},}
where the period matrix in the right-hand side is evaluated
at the attractor point.

The action $S$ has a further nice feature:  Not only it reproduces
the attractor equation by extremization, its value at
the extremum is the semi-classical answer for the entropy!
One can see this since \action\ is quadratic in $\Omega$ and
its extremization gives
\eqn\classicalentropy{ S = {\pi \over 4}
\int \Omega \wedge \bar\Omega_{\bigl|{\rm attractor~value}}=
-{\pi \over 2} {\rm Im}\tau_{IJ} X_{p,q}^I \bar X_{p,q}^J.}
Later in this paper we will be making contact in the semi-classical limit
between this action and the overlap of topological string wave-functions.
Note that the action depends not only on the complex structure moduli
of the Calabi-Yau three-fold but also on the overall scaling factor
$X^I$. We will identify physical degrees of freedom for this extra factor
and how it combines with the complex structure moduli and described
by the action \action .


\newsec{Further Compactification on $S^1$ and Mini-Superspace}

In the context of the black hole entropy counting, it is natural
to consider a further compactification on $S^1$ with periodic
(supersymmetry preserving) boundary conditions on the fermions
around the circle. As we will see, this set-up naturally leads to
computation of the Witten index for black hole microstates in
the Euclidean setup.
Consider type IIB superstring compactified on the Calabi-Yau
3-fold $M$ times $S^2 \times S^1$. There is a natural Euclidean solution 
to the classical equations of motion which develops from this
spacelike section. It is the geometry $M \times S^2 \times
H_2/{\bf Z}$, where $H_2$ is the hyperbolic disk, $i.e.$ the
Euclideanized $AdS_2$, with the metric
\eqn\adsmetric{ ds^2 = d\rho^2 + e^{2\rho} d\tau^2,}
and the ${\bf Z}$ quotient periodically identifies $\tau \sim \tau
+ \beta$, as shown in figure 1. Here we regard $\tau$ as the spacelike
coordinate around the compactified circle $S^1$, and $\rho$ is the
Euclideanized time direction. This geometry is invariant under
supersymmetry, which squares to become a translation along the
$\tau$ direction.

There is another way to view this geometry. A more traditional way
to think about the metric \adsmetric\ is to view $\tau$ as an
Euclideanized time. Since the $\tau$ direction is compactified on
$S^1$ with the supersymmetry preserving boundary condition, the
vacuum amplitude of the string theory in this geometry can be
regarded as the Witten index ${\rm Tr}(-1)^F e^{-\beta H}$, where
$H$ is the translation generator along the $\tau$ direction. The
Witten index counts the degeneracy of ground states of the
supersymmetric black hole discussed in the previous subsection,
and it will play an important role in the following sections.

\bigskip
\centerline{\epsfxsize 3.5truein\epsfbox{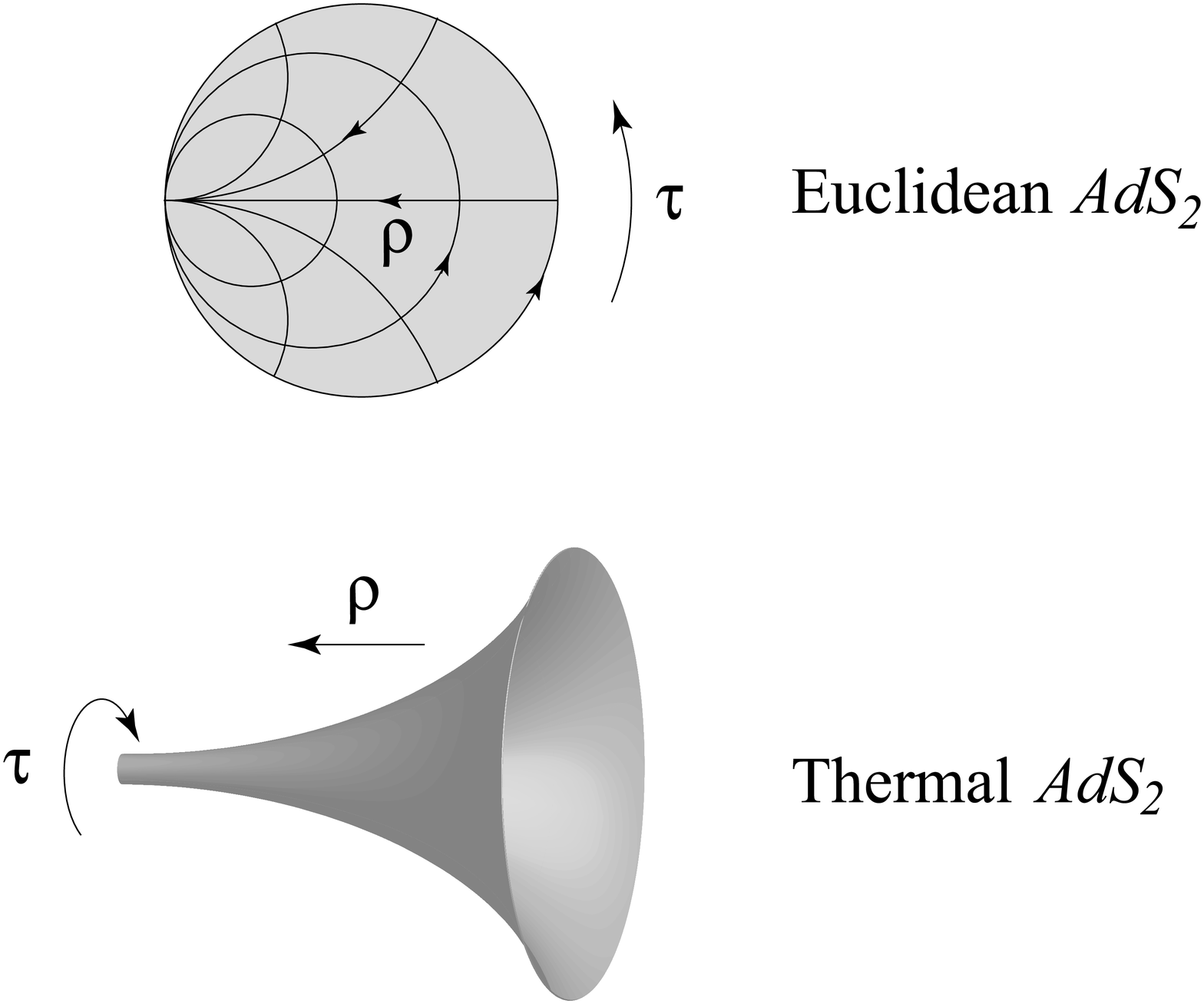}} \leftskip 2pc
\rightskip 2pc \noindent{\ninepoint\sl \baselineskip=2pt {\bf
Fig.1} {{Euclideanized $AdS_2$  is periodically identified in
$\tau$ to make the thermal $AdS_2$ with the supersymmetry
preserving boundary condition. We regard $\rho$ as an Euclidean 
time, which flows from $\rho =+\infty$ to $\rho = -\infty$.}}}
\bigskip

On the other hand, if we view $\rho$ as an Euclideanized time, the
geometry $M \times S^2 \times H_2/{\bf Z}$ describes an Euclidean
time evolution of type II string compactified on $M \times S^2
\times S^1$. This is how we were originally led to the metric
\adsmetric . In this context, it is natural to consider
wave-functions solving the WDW equation on $M \times S^2 \times
S^1$. Since the space is compact, the Calabi-Yau moduli are not
fixed but rather fluctuate, and one should consider a wave
function that is a function of these moduli.

The relation between these two points of view, depending on
whether one regards $\tau$ or $\rho$ as the Euclideanized time in
the target space, is analogous to the worldsheet duality relating
an open string annulus diagram to an exchange of closed string
states. The Witten index computation taking $\tau$ as the
Euclideanized time direction is analogous to the computation in
the open string channel, whereas the wave-function evolving in the
$\rho$ direction is analogous to the boundary state in the closed
string channel.

Let us consider a natural notion of a `mini-superspace' from the
second point of view, where we view $\rho$ as an Euclideanized
time. Among relevant light modes are the complex moduli of
the Calabi-Yau three-fold denoted by  $z^i$ ($i=1,...,h^{2,1}$),
which are in vector multiplets in four dimensions. 
The gravity multiplet also produces some scalar fields upon
compactification on $S^2 \times S^1$. One is the radius $R$ of $S^2$.
Another scalar field is related to how the $S^1$ is fibered
over $S^2$. More specifically, the compactification of
the four-dimensional metric on $S^1$ produces a Kaluza-Klein
gauge field in three dimensions, so we can consider its magnetic
flux through the $S^2$. The additional scalar field $\varphi$
plays a role of the chemical potential for the flux.  
As we will see in section 5, in the WDW equation, the
radius $R$ and the chemical potential $\varphi$ naturally combine
with the complex structure moduli $z$ of the Calabi-Yau to 
make a ``large moduli space'' with coordinates $X^I$
($I=0,1,...,h^{2,1}$). More explicitly, choose any holomorphic
section $X_0^I(z)$ over the complex structure moduli space and
define
\eqn\largemoduli{ X^I = i\ 2 R e^{i\varphi} \ 
 \left({\bar W(X_0)\over K(X_0,\bar X_0) W(X_0)}\right)^{1/2}
X_0^I.}
Note that the right-hand side is invariant under rescaling of
$X_0^I$ and thus gives a well-defined function over the complex
structure moduli space. It is instructive to further note that the
exponentiated K\"ahler potential and the superpotential computed
for $X^I$  have invariant meanings as,
$$\eqalign{ &K(X,\bar X) = -2\ {\rm Im}\tau_{IJ} X^I \bar X^J 
= (2R)^2 , \cr
&W(X)= q_I X^I - p^I F_I = i\ 2R e^{i\varphi} \ M_{BPS}(z,\bar z) .}$$

In four dimensions, 
there are several gauge fields $A^I$ ($I=0,...,h^{2,1}$).
One linear combination of these belongs to the gravity
multiplet, $i.e.$ it is the gravi-photon, and the others are 
in the vector multiplets. 
Upon compactification on $S^1$, each gauge field
becomes equivalent
to a pair of massless scalar fields -- one is the Wilson line 
of the gauge field along the $S^1$, 
and the other is the dual magnetic potential around the $S^1$,
\eqn\whatphi{ \phi^I = \oint_{S^1} A^I, 
~~~ \tilde\phi_I = \oint_{S^1} \tilde{A}_I ,}
where $\tilde A_I$ is the dual of $A^I$ in four dimensions. 
One can also think of $\tilde \phi_I$ as the dual of the
massless gauge field in three dimensions.
  By definition,
they couple to the charges $(p^I, q_I)$ of the
black hole (or the fluxe quantum numbers through $S^2$
 in the near horizon geometry) as
\eqn\chargecoupling{ e^{i \sum_I \left(q_I \phi^I + p^I \tilde 
\phi_I \right)}. }
Combining with $X^I$ defined in \largemoduli, 
s set of four scalar fields $(X^I, \bar X^I, \phi^I, \tilde \phi_I)$
for each $I$ gives bosonic components of supermultiplets.
 
There
are also hypermultiplet fields, which we denote collectively by
$H$, which consist of the K\"ahler moduli of the Calabi-Yau
together with the RR-forms. We should also consider the radius 
$\beta$ of the $S^1$. A natural
vacuum wave-function will then depend on these variables:
$$\Phi_{p,q}(X^I, \bar X^I; \phi^I, \tilde\phi_I; H, \beta).$$
We also noted the dependence of the wave-function
on the discrete choice of the 5-form
flux $F_5$ captured by the set of integers $(p^I,q_J)$. 

For the vacuum state we would expect that the wave-function should
not depend on the hypermultiplets $H$ or the radius $\beta$, 
at least to all orders in string perturbation theory. 
Since
these moduli are not fixed by the classical equations of motion
and also to all order in the perturbative expansion, it is natural
to expect that the ground state wave-function does not depend on
them. 

The dependence of the wave-function on $(\phi^I, \tilde\phi_I)$ is
also simple. Since $\phi^I$ is dual to $\tilde A_I$ and
$\tilde \phi_I$ is dual to $A^I$ in three dimensions, we have
\eqn\duality{\eqalign{
&G_{IJ} {d\phi^J \over d\rho}
 = \int_{S^2} \tilde F_I = q_I, \cr
&G^{IJ} {d\tilde\phi_J \over d\rho}
 = \int_{S^2} F^I = p^I,}}
where $G_{IJ}$ is the metric in the kinetic term for the gauge
fields, and $F^I = dA^I$, $\tilde F_I = d\tilde A_I$.
These equations means that, when we quantize the theory along 
the $\rho$ direction, $(\phi^I, \tilde \phi_I)$
are canonically conjugate to $(q_I, p^I)$. Therefore, the wave-function
depends on $(\phi, \tilde\phi)$ as
\eqn\phidependence{
  \Phi_{p,q}(X, \bar X; \phi, \tilde \phi)
 = e^{i\sum_I (q_I \phi^I + p^I \tilde \phi_I)}
\Psi_{p,q}(X, \bar X),}
if it is an eigenstate
of the fluxe quantum numbers $(p,q)$. This $(\phi, \tilde\phi)$ dependence
is also expected from the fact that $(\phi, \tilde\phi)$
are electric and magnetic static potentials for the black
hole charges \chargecoupling . 

Thus, it is natural to look for a wave-function $\Psi$ which
depends only on the $X^I$ and the amount of fluxes determined by
$p,q$,
$$\Psi_{p,q}(X,{\bar X}).$$
This wave-function should have the property that $|\Psi_{p,q}|^2$
is strongly peaked at the attractor value of the moduli
\attraction\ which is what extremizes the superpotential.

Do we know anything a priori about this mini-superspace
wave-function? One question we may ask for such a wave-function is
whether we can compare the different values of fluxes $(p,q)$ or
if that is a super-selection sector. Since the dimensional reduction
of the four-dimensional vector multiplets gives rise to 
the electric and magnetic potentials $(\phi^I, \tilde\phi_I)$
as dynamical variables and since they are conjugate to
$(q,p)$, the wave-function $\Psi_{p,q}(X,\bar X)$
should be defined over the entire landscape of vacua with different
values of $(p,q)$. In fact, since we do not have any more
spatial dimensions left, changing the fluxes does not cost
infinite action and we should be able to compare different values
$(p,q)$. This is not just a matter of principle, but in fact there
are instantons that interpolate different amounts of fluxes. Such
an instanton can be constructed as a domain wall D3 brane that
wraps 3-cycles in the Calabi-Yau three-fold and winds around the
$S^1$. Its action is $|\delta W|$ times $\beta$, where
$\delta W$ is the variation of the superpotential \gvwpotential\
corresponding to the change of the fluxes.

In such a situation, would it be reasonable to expect the
probability of the wave-function for the different flux sectors to
be the same, namely
$$\int dXd{\overline X} \ \left| \Psi_{p,q}(X,\bar X)\right|^2=1 ~ ?$$
This may sound like the most economical assumption, in that we
would be weighing each discrete choice of flux vacua equally.
However, there is a reason to think that this is not natural.
Consider the original Hartle-Hawking wave-function for a
three-sphere $S^3$. The saddle point computation of the
wave-function on $S^3$ can be viewed as filling it with a
$4$-dimensional ball with the $S^3$ as its boundary, and this
leads to the action $S_E$ in the Euclidean ball,
$$S_E \sim - {1\over \Lambda},$$
where $\Lambda$ is the cosmological constant, and the
mini-superspace wave-function behaves as
$$\Psi \sim \exp\left(-{1\over \Lambda}\right).$$
In the present context $\Lambda<0$, and $- {1\over \Lambda} \sim
{\rm Area}(S^2) \sim S_{\rm entropy}$, so we may expect that
\eqn\psisquared{\int dXd{\overline X}\  \left| \Psi_{p,q}(X,\bar
X)\right|^2 \sim \exp\left(S_{\rm entropy}\right),}
namely the wave-function is normalized by the exponential of the
entropy. It is natural since the string partition function on the
full space $M \times S^2 \times H_2 /{\bf Z}$ should give the black hole
entropy. Thus, at least semi-classically we expect \psisquared\ to
hold.

In view of our discussion following \action\ a natural guess for
the probability measure is
\eqn\semiclassicalweight{|\Psi_{p,q}(X,{\bar X})|^2
\sim \exp\left[-{\pi\over
4} K-{\pi \over 2}i\left(  W-{\bar
W}\right)\right],}
where $W(X)$ depends on the flux determined by
$(p,q)$ as in \bpsmass. Indeed, the right-hand side is peaked at the
attractor value and its value is given by the exponential of the
entropy. To see this in terms of the physical variables, the
complex structure moduli $z^i$ and the radius $R$ of the $S^2$, we
can substitute \largemoduli\ into \semiclassicalweight\ and find
\eqn\physvariables{ |\Psi_{p,q}|^2
\sim\exp\left[ -\pi\left( R^2 -  M_{BPS} R\right)\right].}
Extremizing this with respect to the complex structure moduli
$z^i$ gives the attractor equation as we discussed in section 2.1
and extremizing with respect to $R$ gives
$$ |\Psi_{p,q}|^2_{~~\bigl|{\rm extremum}}  \sim
\exp\left({\pi \over 4} M_{BPS}^2\right) = \exp(S_{\rm
entropy}),$$
reproducing the expected result.

In section 5, we will
study a semi-classical function that solves the WDW equation.
We will find \semiclassicalweight\ captures essential aspects
of the wave-function. This same wave function will give a
semi-classical approximation to the topological string partition
function, which we will argue gives the exact answer for the
Hartle-Hawking wave function including all string loop
corrections.

Note that  \semiclassicalweight\ suggests
that we can
write the flux dependent part of the wave function as either
a holomorphic function of $X$, given by
$$\exp\left[-i{\pi\over 2} W_{p,q}(X)\right],$$
(where we have written the subscripts of $(p,q)$ for $W$
to recall that $W$ depends on them),
or in a real basis
$$\exp\left[-i{ \pi\over 4}\left( W_{p,q}(X)- \bar W_{p,q}
(\bar X)\right) \right].$$
(A more accurate description of this will be given in
sections 5 and 6.)
In either case this suggests that\foot{For this description,
we should define $\Psi_{p,q}$ from a state $\Psi$
in a bigger Hilbert space including
the choice of fluxes so that
$$\langle p,q|\Psi \rangle \rangle =|\Psi_{p,q}\rangle .$$
}
we should view the $p,q$ dependence of
the state as obtained by an operator $O_{p,q}$ acting on the state
corresponding to $p,q=0$
$$|\Psi_{p,q}\rangle =O_{p,q}|\Psi_{0,0}\rangle .$$
Note that we also find that semi-classically
$\Psi_{0,0}(X,{\overline X})=1$. It is a nice feature that this is
an affine space in the sense that
$O_{p,q}.O_{p',q'}=O_{p+p',q+q'}$.

{}From this heuristic form of the wave function we see an extra
unexpected structure which will play a key role in the following:
The wave function is essentially a function of half of the variables
(in the holomorphic description only a function of $X$ but not ${\overline X}$).
As we will discuss below this is a reflection of the fact that we are considering
the mini-superspace relevant for BPS quantities.  
This implies that on the reduced phase space of BPS quantities
$X$ and ${\overline X}$ do not commute so that the wave
function should have been only a function of half the variables.
This also gives  another reason why the wave-function does not
depend on the hypermultiplet fields $H$ and the radius $\beta$ of
$S^1$ as well as massive string fields.
In order to show this we first turn to a discussion of attractor flow
which is the equation of motion for the BPS quantities on the mini-superspace.

\newsec{Attractor Flow Equations}

The wave-function $\Psi_{p,q}(X,\bar X)$ should satisfy the
Wheeler-De Witt equation.  In mini-superspace the WDW equation
corresponds to the quantization of the attractor flow for a black
hole with charges $p^I$ and $q_I$. Therefore, in this section we
review the equations that govern the classical attractor flow
\refs{\fks,\asm,\FGK,\denef}.  The flow parameter
will be identified with the Euclidean time $\rho$ in our setup.

Consider a ten-dimensional Euclidean metric of the form,
\eqn\tendmetric{
 ds^2 =  e^{2U+2\rho}  d\tau^2 + e^{-2U}d\rho^2+ e^{-2U} d\Omega^2
+ ds^2_{\rm CY}, }
where $\tau$ is the Euclideanized time direction compactified on
$S^1$, $\rho$ is the radius coordinate,
$d\Omega^2$ is the metric on a two-sphere
of unit radius, and $ds^2_{\rm CY}$ is the metric on the internal
Calabi-Yau three-fold. Note that $e^{-U}$ is the radius of the
$S^2$, and the $AdS_2$ geometry is realized when $U$ is constant.
Since we are interested in BPS configurations and since the
supercharges preserved by the background square to become
the translation along the $\tau$ direction, we assume that 
the scale factor $e^{U}$ and the complex
moduli moduli $z^i$ ($i=1,...,h^{2,1}$) of the Calabi-Yau
three-fold are independent of $\tau$. In this case, we have
a one-dimensional system along the $\rho$ direction described
by the effective action  \FGK ,
\eqn\eaction{
\eqalign{S_{{\rm eff}} = {1\over 2} \int_{-\infty}^\infty d\rho
e^\rho &\left[ \left({dU \over d\rho}+1\right)^2  + g_{i\bar j} {dz^i
\over d\rho} {d\bar z^{\bar j} \over d\rho} \right. \cr
&\left. + e^{2U}\left(  M_{BPS}^2(z,\bar z)
+ 4 g^{i\bar j} \partial_i M_{BPS} \bar\partial_{\bar j}
M_{BPS} \right)\right],}}
where
\eqn\whatmbps{ M_{BPS} = \sqrt{ {W(z) \bar{W}(\bar z)\over
K(z,\bar z)}} .}
We regard $\rho$ as the Euclidean time of the system,
which flows from $\rho = +\infty$ to $-\infty$. 
Since the effective action \eaction\ can be written as
$$
\eqalign{ S_{{\rm eff}} = {1\over 2} 
\int_{-\infty}^\infty d\rho e^\rho &\left[
\left( {dU \over d\rho} +1 - e^U M_{BPS}\right)^2
\right.
 +   \cr
&~~ \left.  + g_{i\bar j}\left( {dz^i \over d\rho} - 2e^U g^{i\bar m}
\bar{\partial}_{\bar m} M_{BPS}\right) \left( {d\bar z^{\bar j}
\over d\rho} - 2 e^U g^{\bar j n}
\partial_n M_{BPS} \right) \right]
\cr
& ~~+ ({\rm total~derivative}),}$$
the BPS equations are
\eqn\standardbps{ \eqalign{ {dU \over d\rho} & = - 1 + e^U
M_{BPS}(z,\bar z), \cr
         {dz^i \over d\rho} & = 2 e^U g^{i\bar j}
\bar{\partial}_{\bar j} M_{BPS}(z,\bar z) .}}
The signs on the right-hand side of these equations are chosen 
so that they are compatible with the initial condition at 
$\rho \rightarrow \infty$, which we regard as the infinite
past in the Euclidean time.

The equations \standardbps\ can be combined into a single equation
on the large moduli space. To write down such an equation, we
start with a holomorphic section $X_0^I(z)$ ($I=0,1,...,h^{2,1}$)
over the moduli space of complex structure. They make projective
coordinates of the moduli space, and as such there is a freedom to
rescale these coordinates. We define the exponentiated K\"ahler
potential $K_0$ and the superpotential $W_0$ for these coordinates
as
$$ \eqalign{ &K_0  = -2\ {\rm Im}\tau_{IJ} ~ X_0^I \bar X_0^J , \cr
   & W_0 = q_I X_0^I - p^I F_I(X_0). }$$
We then combine the scale factor $e^U$ in the metric \tendmetric\
and the complex moduli $z^i$ into a single set of coordinates
$X^I$ defined by
\eqn\combine{
   X^I = 2i e^{-U} \left( {\bar W_0 \over K_0 W_0} \right)^{1/2}
  X_0^I.}
(This is the same as \largemoduli\ with the identification
that $e^{-U}$ is the radius of the $S^2$. Note that we are setting
$\varphi=0$ since the $S^1$ is trivially fibered over $S^2$
in \tendmetric .)
Note that the right-hand side of \combine\ is invariant under
rescaling of $X_0^I$. Moreover
$$K(X,\bar X)= -2 \ {\rm Im}\tau_{IJ} X^I \bar X^J
=\left(2 e^{-U}\right)^2$$
is the diameter squared of the $S^2$. Thus, the large moduli space
parametrized by $X^I$ combines the complex moduli $z^i$ and the
radius of the $S^2$, as we suggested earlier in section 3.  Using
$X^I$, the attractor flow equations \standardbps\ can be written
as a single equation
\eqn\singlebps{ {dX^I \over d\rho} =   X^I + \left({i \over {\rm
Im}\tau}\right)^{IJ} \bar \partial_J \bar W(\bar X). }

Let us show that \singlebps\ is equivalent to \standardbps . If we
multiply $ \bar X^J {\rm Im}\tau_{JI}$ to both sides of
\singlebps, the left-hand side becomes
$$ \eqalign{ &  \bar X^I {\rm Im}\tau_{IJ}{d X^J \over d\rho} \cr
& = 2e^{-2U} {dU \over d\rho}
    - {e^{-2U}\over M_{BPS}}\left( {dz^i \over d\rho} \partial_i M_{BPS}
                 - {d\bar z^{\bar i} \over d\rho} \bar\partial_{\bar i}
                    M_{BPS} \right) .}$$
On the other hand, the right-hand side becomes
$$  \eqalign{  X^I {\rm Im} \tau_{IJ} \bar X^J + i \bar X^I
\partial_I \bar W ( \bar X_0)
 & = -{1\over 2}K(X,\bar X)+  i \bar W (\bar X) \cr
&= -2 e^{-2U} + 2 e^{-U}M_{BPS}.} $$
Combining them together, we obtain
$$ {dU \over d\rho} - {1 \over 2M_{BPS}} \left( {dz^i \over d\rho} \partial_i M_{BPS}
                 - {d\bar z^{\bar i} \over d\rho} \bar\partial_{\bar i}
                    M_{BPS} \right)
= - 1+ e^{U} M_{BPS}. $$
The real part of this equation is precisely the first of
\standardbps : \eqn\first{ {dU \over d\rho} = -1+ e^U M_{BPS}.}
 The imaginary part gives
\eqn\additional{ {dz^i \over d\rho} \partial_i M_{BPS} = {d\bar
z^{\bar i} \over d\rho} \bar\partial_{\bar i} M_{BPS} .}
Similarly multiplying $\partial_{i} X_0^J {\rm Im}\tau_{IJ}$ to
both sides of \singlebps\ and using \first\ and \additional , we
find
\eqn\second{ {dz^i \over d\rho} = 2e^U g^{i \bar j}
\bar\partial_{\bar j} M_{BPS}.}
Moreover, \second\ implies \additional\ since both sides of
\additional\ are now equal to $-2 e^U g^{i\bar j} \partial_i
M_{BPS} \bar\partial_{\bar j} M_{BPS}$. Therefore, \singlebps\ for
$X^I$ defined by \combine\ is equivalent to the standard BPS
equations \standardbps .

In \singlebps, a general BPS solution can be easily expressed.
Taking the real and imaginary parts of this equation, one finds
$$  {\rm Re}\left(X^I-{dX^I\over d\rho} \right) = p^I,
~~~~ {\rm Re}\left(F_I- {dF_I\over d\rho}  \right) = q_I.$$ A general
solution to this is then
$$\eqalign{ & {\rm Re} X^I=p^I + c^I e^{\rho}, \cr
& {\rm Re} F_I=q_I + d_I e^{\rho},}$$
where $(c^I, d_I)$ are integration constants specified by the
initial condition at the infinite past $\rho=\infty$. Whatever
initial condition one chooses there, $X^I$ at the
infinite future $\rho\rightarrow
-\infty$ are fixed to be at the attractor value,
$$ {\rm Re}(X^I) \rightarrow p^I,~~
{\rm Re}(F_I) \rightarrow q_I.$$

It is useful to write the BPS
equation \singlebps\ as
\eqn\classicalbps{ {\rm Im}(\tau_{IJ}) {d X^I \over d\rho} =-{1\over
2} \bar \partial_I K + i \partial_I \bar W.}
In the next section we will use this equation to obtain the
supersymmetric version of the WDW equation. Namely, after
quantization it gives the BPS condition that when imposed on the
states implies the WDW equation.  In order to do this it is useful to keep in mind
that the above BPS condition \classicalbps\ can be derived from a local action
with 4 supercharges of the form 
$$\int d^4\theta\  K +[\int d^2\theta \ W+c.c.].$$
The fact that the derivatives are replaced by covariant derivatives in \classicalbps\
simply reflects the fact that we are discussing a local (gravitational)
theory with 4 supercharges.
This will be used in the next section to determine the commutation relation
of fields restricted to the BPS sector.

\newsec{Wheeler-De Witt Equation and the Semi-Classical Wave Function}

In this section, we describe the semi-classical Hartle-Hawking
type wave function for type II superstring associated to the
spacelike geometry $M\times S^2 \times S^1$.

\subsec{Wheeler-De~Witt Equation in Mini-Superspace}

We recall some basic aspects of WDW equation in the
mini-superspace approach. The WDW equation is in general an
infinite dimensional wave equation which takes the form,
$$H|\Psi \rangle =0,$$
where $|\Psi \rangle$ is a function of the {\it superspace},
which is the space of all possible metric data
on the spacelike hypersurface, and $H$ is the Hamiltonian 
for the gravitational system. In the mini-superspace 
approach, $|\Psi \rangle$ becomes a
wave function on a finite dimensional subspace of the superspace. 
In particular the above equation becomes similar to a ordinary
wave equation for a quantum mechanical system.  Most of the 
subtlety in solving such a reduced 
system is hidden in the choice of boundary conditions in solving
this quantum mechanical system.

In the case at hand we have a stronger version of WDW equation,
because we have supersymmetry.  More precisely we are looking for
a state $|\Psi \rangle$ which is supersymmetric,
$$Q|\Psi >=0, $$
with $Q^2=H$. Moreover, in this case, the reduction to the
mini-superspace is not an approximation but is an exact
description of the wave-function for computations involving
BPS quantities.

The corresponding quantum mechanical system  is a
theory with 4 supercharges, with a superpotential
$W=q_IX^I-p^IF_I$. This situation is a local (gravitational)
version of the more familiar global (non-gravitational) LG
theories with 4 supercharges, which we will now review, in order
to gain insight into the case at hand. 
In particular we will
consider the case of conformal theories where the corresponding
superpotential $W(\Phi^i)$ is a quasi-homogeneous function of
chiral fields $\Phi^i$.

\bigskip

\bigskip
\noindent $\circ$ {\it Global (Non-Gravitational) Case}
\medskip

This theory has been studied both in $(1+1)$ dimensions as well as
in $(0+1)$ dimension, and as far as the vacuum geometry is
concerned it is independent of the dimension, since the modes
reduce to constant modes. As discussed in \ref\ttstar{S.~Cecotti
and C.~Vafa, ``Topological anti-topological fusion,'' Nucl.\
Phys.\ B {\bf 367} (1991) 359.
}\ one can define a canonical notion of a distinguished vacuum
state $|0\rangle$. This is done more naturally in the context of
the $(1+1)$ dimensional theory where one consider the
topologically twisted theory on a semi-infinite cigar, where the
space is identified with the boundary circle.  This gives a
holomorphic basis of the vacuum.  In particular $|0\rangle$ does
not depend on the coupling constants of $\bar W$.  With this
choice of the vacuum state, the normalization of the vacuum state
is unambiguous, and the norm
$$\langle {\bar 0}|0\rangle$$
is well-defined.  In fact, in the conformal case,
it can be shown \ttstar\
that this overlap is given by
\eqn\overl{\langle {\bar 0}|0\rangle =\int
d\Phi d\bar \Phi \ {\rm exp}(-W(\Phi)+W(\bar \Phi)).}
Moreover there is a sense in which $\exp(-W(\Phi))$ is the wave function
associated to $|0\rangle$ or more precisely in
the same cohomology.   In particular correlation functions of
chiral operators evaluated in the vacuum are given by integrals involving
$\exp(-W(\Phi))$ as
$$\langle \bar 0|f(\bar \Phi) g(\Phi)|0\rangle =\int d\Phi d\bar \Phi
\ f(\bar \Phi) g(\Phi) {\rm exp}(-W(\Phi)+W(\bar \Phi)).$$
Another way to state it is   \lref\hiv{K.~Hori, A.~Iqbal and C.~Vafa,
``D-branes and mirror symmetry,''
{\tt hep-th/0005247}.
}\
\refs{\ttstar,\hiv}
$$\langle \gamma|0\rangle =\int_\gamma d\Phi \ {\rm exp}(-W(\Phi)),$$
where $\langle \gamma |$ is a boundary state defined by the Lagrangian D-brane
related to $\gamma$. Given this, \overl\ follows since
$|\gamma\rangle$ form a basis for ground states and
we can use the Riemann bilinear identity to evaluate integrals.

A quick/heuristic way to see the relation between
$|0\rangle$ and ${\rm exp}(-W(\Phi))$
is to note that since
$$Q|0\rangle=0, $$
the supersymmetric flow conditions must be satisfied when
evaluated on $|0\rangle$.  The supersymmetric gradient flow is given by
$$ G_{\bar i j}{d\bar \Phi^i\over d\rho}
-{\partial W\over \partial\Phi_j}=0.$$
Upon quantization, we have
$$G_{\bar i j}{d\bar \Phi^i\over d\rho}\rightarrow - {\partial \over
\partial\Phi_j},$$
(note that we do not have $(-i)$ but $(-1)$ in the right-hand side
since $\tau$ is the imaginary time)
and the flow equation becomes
\eqn\globalflow{ \left( {\partial \over \partial \Phi^i} +
{\partial W\over\partial \Phi^i} \right) \psi = 0, }
and one arrives at $\psi(\Phi^i)\sim {\rm exp}(-W(\Phi^i))$. This
is a rough derivation, and one could consult \refs{\ttstar, \hiv}
for more precise statements.

\bigskip
\noindent
$\circ$ {\it Local (Gravitational) Case}
\medskip

Now we turn to the case of interest for this paper.  This differs
in two important respects relative to what we just discussed.
First of all, it is a local theory, involving gravity in addition
to matter. A second difference is that the expression
$$W(X)=q_I X^I- p^I F_I$$
is only a leading computation of the superpotential.  In
particular one would expect that $F_I$ receives string quantum
corrections as is suggested by the topological string amplitudes.

At first sight there is a third difference. Namely, in our
situation the state $|\Psi\rangle$ is represented by a wave
function $\Psi_{p,q}(X,\bar X)$ on the full configuration space,
and not (yet) by a holomorphic function $\psi_{p,q}(X)$. But, as
we will explain below, this difference will disappear once we make
use of the WDW equation. The supersymmetric WDW equation is
equivalent to the quantum version of the BPS equation \singlebps\
and its complex conjugate. To understand the quantum version of 
the BPS equation, we note that the metric in the $X$ space
implied by the effective action \eaction\ is ``almost'' given
by ${\rm Im}(\tau_{IJ})$ since
\eqn\almostmetric{\eqalign{
  &X^I {\rm Im}( \tau_{IJ}) \bar X^J = - 2R^2, \cr
  & D_i X^I {\rm Im}( \tau_{IJ})  \bar X^J
 = X^I {\rm Im}( \tau_{IJ}) \bar D_{\bar j} \bar X^J = 0\cr
&D_i X^I {\rm Im}( \tau_{IJ})  \bar D_{\bar j} \bar X^J =
2R^2 \ g_{i\bar j},}}
where
$$ D_i X^I = K \partial_i \left( K^{-1} X^I\right),~~
   \bar D_{\bar j} \bar X^J  = K \bar\partial_{\bar j}
     \left( K^{-1} \bar X^J\right) , $$
and $g_{i\bar j} = \partial_i \bar \partial_{\bar j} \ln K$.
We note that ${\rm Im}\ \tau_{IJ}$ has one negative sign
in the direction of $H^{(3,0)}$ whereas it gives the standard
positive definite metric in the $H^{(2,1)}$ direction. Flipping
the sign in the $H^{(3,0)}$ direction gives what is denoted
by ${\rm Im}\ {\cal N}_{IJ}$ in the supergravity literature,
which is the metric derived from the effective action \eaction.
In the semi-classical approximation,
flipping of the sign of the metric can be done by a suitable
contour deformation in a functional integral, which we will
freely do in this section. Thus, we will use ${\rm Im}\ \tau_{IJ}$
as our metric and the corresponding quantization rule is 
$$
{\pi \over 4} {\rm Im}( \tau_{IJ}) {d X^J\over d\rho} \rightarrow
- {\partial \over
\partial \bar X_J}.$$
(We did not derive the factor $\pi/4$ in the right-hand side, but
it was introduced in order for the quantum BPS equation to be consistent
with the black hole entropy computation.)

Given this rule, the quantum version of the BPS equation \singlebps\ is 
\eqn\localflow{\left({\partial\over \partial \bar X^I}- {\pi \over
8} \bar \partial_I K +i{\pi \over 4}\bar\partial_I \bar
W\right)\Psi_{p,q}=0,}
and similarly for the complex conjugate equation
\eqn\localflows{ \left({\partial\over \partial X^I}- {\pi \over 8}
 \partial_I K -i{\pi \over 4}\partial_I
 W\right)\Psi_{p,q}=0.}
For future reference, let us denote the operator appearing in the constraint \localflow\ 
$\bar C_I$ and
the one in \localflows\ by $C_I$, 
so that in terms of the state $|\Psi_{p,q}\rangle$ the constraints are 
$\bar C_I|\Psi_{p,q}\rangle=0$ and $C_I|\Psi_{p,q}\rangle=0$ respectively.
Imposing both constraints is sufficient to determine the entire
wave function. One finds in this way
$$
\Psi_{p,q}(X,\bar{X})=\exp\left[ {\pi\over
8}K(X,\bar{X})+{\pi \over 4}\left(iW(X)-i\bar{W}(\bar
{X})\right)\right].
$$

One can use the BPS constraints also in a way that leads to a
description of the covariant supersymmetric gradient flow, also
known as the attractor flow, in terms of a holomorphic wave
function $\psi_{p,q}(X)$. This wave function can be 
obtained by imposing first the constraint \localflow. This reduces
$\Psi(X,\bar{X})$ essentially to a holomorphic function. The
second condition \localflows\ then fixes $\psi_{p,q}(X)$ to be
given by
\eqn\wavy{\psi_{p,q}(X)=e^{i{\pi\over 2}W_{p,q}(X)}.}
(See section 5.2 for further discussion on the holomorphic
wave function.)

Needless to say this derivation is heuristic, but nevertheless, it
is expected that at the semi-classical level $\psi_{p,q}$ should
give a good approximation to the wave function.  Note in
particular that $\psi_{0,0}(X)=1$.  In the remaining part of this
paper we will provide evidence that this wave function coincides
with the semi-classical approximation obtained from the topological
string partition function, which we propose to be the exact
Hartle-Hawking wave function including all string loop
corrections.

\subsec{Reduced phase space, and the correspondence with
topological strings}

There is an alternative way to view the result \wavy. Instead of
imposing the quantum BPS condition \localflow , one can first
solve the BPS condition \classicalbps\ on the classical variables
and then quantize the reduced system. The solutions of the first
order BPS equations are determined by the initial values for $X^I$
and $\bar{X}^I$, and hence these parametrize the phase space of
BPS configurations. In fact, the BPS equations can be viewed as
second class constraints on the original phase space variables
that include also the time derivatives $dX^I/ d\rho$ and
$d\bar{X}^I/ d\rho$. The original Poisson brackets then
change in to the Dirac brackets, and as a result $X^I$ and
$\bar{X}^I$ become non-commuting variables.

We abbreviate the operators appearing in the constraints
\localflow\ and \localflows\ by $\bar C_I$ and $C_I$ respectively, as mentioned above. 
We want to impose $\bar{C}_I$ on the ket state $|\Psi\rangle$ and its
conjugate $\bar{C}^\dagger_J$ on the bra state $\langle\Psi|$.
Notice that $\bar{C}^\dagger_J$ differs from $C_J$ in the sign of
the derivative.  The Dirac bracket is defined as
$$
\Bigl[ X^I, \bar X^J\Bigr]_{{\rm Dirac}} =
\Bigl[X^I, \bar X^J\Bigr] - \sum_{K,L} \Bigl[ X^I, \bar C_K^\dagger \Bigr]
{1\over  \Bigl[\bar{C}_K^\dagger ,\bar{C}_L\Bigr]}
\Bigl[\bar C_L,\bar X^J\Bigr], 
$$
where the denominator should be read as the inverse matrix. Here
we only wrote commutators that we know are non-vanishing. For the
constraints one finds
$$
\Bigl[ \bar{C}_K,\bar{C}^\dagger_L\Bigr]=-{\pi\over
4}\bar\del_K\del_L K= {\pi\over 2}{\rm Im}\tau_{KL},
$$
while the commutators of the constraints with the coordinates give
$$
\Bigl[X^I,\bar C_K^\dagger \Bigr] =\delta^I{}_K,\qquad\quad
\Bigl[\bar C_L,\bar X^J_{\,}\Bigr]=\delta_L{}^J.
$$
Inserting this in to the definition for the Dirac bracket leads to
the following commutation relations for $X^I$ and $\bar{X}^J$
\eqn\noncommutative{ \Bigl[ X^I, \bar X^J\Bigr]_{{\rm Dirac}} 
= {2 \over \pi}
\left({1\over {\rm Im}\tau}\right)^{IJ}.}
We like to stress that the non-commutative nature of the $X^I$ and
$\bar{X}^I$ only arises because we are looking at the BPS
configurations, and does not hold in the full theory. Physically
what this means is that if we used the measuring devices made {\it
only} of BPS objects we cannot simultaneously measure $X^I$ and
$\bar{X}^I$. This situation is analogous to the quantization of
two-dimensional electrons in a constant magnetic field, where the
two-dimensional plane becomes non-commutative when the system is
constrained to be on the first Landau level.  Our derivation of
the canonical commutation relation precisely mimics the analogous
calculation for the Landau problem. The operators $\bar{C}$ and
$\bar{C}^\dagger$ are like the creation/annihilation operators
that appear in the Hamiltonian, while $C$ and $C^\dagger$ are the
creation/annihilation operators that commute with the Hamiltonian
and therefore act on the states in the lowest Landau level. On the constrained
phase space the latter are linearly related to the coordinates $X$
and $\bar{X}$ on the original configuration space.

The commutation relations \noncommutative\ imply that the
symplectic form on the reduced phase space is given by the
K\"ahler form derived from $K(X,\bar{X})$. In geometric
quantization one starts from a phase space with a symplectic form
and then one constructs the `pre-quantum' Hilbert space of wave
functions $\Psi(X,\bar{X})$. Next one picks a polarization and
removes the dependence on half of the coordinates. For a
K\"{a}hler manifold the natural polarization is to impose that
wave functions are holomorphic. The Hilbert space for
\noncommutative\ can be thus be represented by holomorphic
wave-functions $\psi(X^I)$ of $X^I$, with the inner product
defined by
\eqn\reducedmetric{
\langle \psi_1 | \psi_2\rangle =\int dX d\bar X \ e^{{\pi \over 4} K}
\bar\psi_1(\bar X) \psi_2(X).}
Note that the weight factor $e^{{\pi \over 4}K}$ is
determined so that the commutation relation \noncommutative\
is compatible with the fact that $X^I$ and $\bar X^J$ are
hermitian conjugate in the reduced Hilbert space. 
The constraint \localflow\ is identical to the restriction imposed
in K\"{a}hler quantization on the pre-quantum wave functions
$\Psi(X,\bar{X})$ to get the quantum wave functions $\psi(X)$.

So far we have used the constraints $\bar C_I, \bar C_I^\dagger$. 
The other set of the constraints $C_I, C_I^\dagger$ determined
the holomorphic wave-function as in \wavy .
Its norm square according to the inner product \reducedmetric\
is then
$$ \langle \psi_{p,q} | \psi_{p,q} \rangle
= \int dX d\bar X 
    \exp\left[{\pi \over 4} K(X,\bar X)
+{\pi \over 2}i \left(W(X) + W(\bar X)\right)\right].$$
By performing the analytic continuation $(X,\bar X)
\rightarrow (-X,\bar X)$, this becomes
\eqn\scnorm{ \langle \psi_{p,q} | \psi_{p,q} \rangle
= \int dX d\bar X 
    \exp\left[-{\pi \over 4} K(X,\bar X)
-{\pi \over 2}i \left(W(X) - W(\bar X)\right)\right].}
This reproduces the semi-classical probability measure
\semiclassicalweight\ we conjectured based on the entropy
argument. After the analytic continuation, the holomorphic
wave function becomes
$$ \psi_{p,q}(X) = e^{-iW(X)},$$
which also agrees with what we expected in section 3. 

The K\"{a}hler quantization of the reduced phase space using the
K\"{a}hler potential $K(X,\bar{X})$ is only possible in the
semi-classical approximation. Due to the non-linearity of
$K(X,\bar{X})$ there are many normal ordering problems that are
hard if not impossible to solve. Fortunately, the quantization of
the reduced phase space can be simplified by making an appropriate
change of variables. Motivated by the form of the attractor
equations, let us introduce real coordinates $\chi^I$ and $\eta_I$
via
\eqn\canonicaltransformation{\chi^I={\rm Re} \left(X^I\right)
\qquad \eta_I={\rm Re} \left(F_I\right).}
To determine the symplectic form for these new coordinates, let us
calculate the commutator of the expressions on the right-hand side at the
level of Poisson brackets, {\it i.e.} ignoring any operator
ordering. One finds
$$
\eqalign{\left[ {\rm Re} \left(X^I\right),{\rm Re}
\left(F_J\right)\right]_{{\rm Dirac}} & = {1\over 4}\left[ X^I,
\bar{F}_J\right]_{{\rm Dirac}}
+ {1\over 4}\left[ \bar X^I, F_J\right]_{{\rm Dirac}}\cr &
={1\over 2\pi} \left({1\over {\rm Im}\tau}\right)^{IK}
\left(\bar{\tau}_{KJ}-\tau_{KJ}\right)\cr &={1\over \pi i
}\delta^I{}_J}
$$
and hence we conclude in the real $(\chi,\eta)$ coordinates the
symplectic form is simply the canonical one leading to standard
commutation relations
\eqn\realcommutator{ \left[ \chi^I, \eta_J\right]_{{\rm Dirac}} = {1 \over \pi
i} \delta^I{}_J.}
The change of variables that we just described has a natural
interpretation in terms of the space of close three forms
$H^3(M,R)$. A three-form $\gamma\in H^3(M)$  can be parametrized
in terms of the coordinates $\chi^I$ and $\eta_J$ as
$$
\gamma=\chi^I\alpha_I+\eta_I\beta^I.
$$
in terms of the canonical basis $\alpha_I$ and $\beta^J$. The
complex coordinates $X^I$ that are used in the WDW equation
represent the periods of the (anti-)holomorphic three form
$\Omega(X)$, which can be written in the canonical basis as
$$
\Omega(X)=X^I\alpha_I+\del_I F_0(X)\beta^I .
$$
The form of the canonical transformation \canonicaltransformation\
follows from by making the identification $\gamma={\rm Re}\ \Omega$.  The
fact that $(\chi,\eta)$ are canonically conjugate variables also
follows from the natural symplectic structure on $H^3$.

What we have achieved now is that the problem of quantizing the
reduced phase space of BPS coordinates $X^I$ and $\bar{X}^I$ is
mapped on to the much simpler problem of quantizing $H^3(M)$ using
the real $(\chi,\eta)$-coordinates. Wave functions are in this
case simply given as $\psi(\chi)$ with the familiar inner product.
The price that one has to pay, however, is that wave function that
are simple in terms of the $X^I$ variables will look complicated
in terms of $\chi$. In particular, we argued that the holomorphic
wave function $\psi_{0,0}(X)$ to be used in the  solution of the
WDW equation for $p=q=0$ is just $\psi_{0,0}(X)=1$. The corresponding wave
function $\psi_{0,0}(\chi)$ in the real variables will not be as
simple, in fact, we will show that it is equal to the topological
string partition function!

The relation between the wave function in the real and complex
polarization can be found at the semi-classical level by applying
standard canonical transformation techniques (see $e.g.$
\goldstein).  In classical mechanics canonical transformations can
be described with the help of a generating function. In our case
this generating function should depend on one of the real and one
of the complex coordinates. Since we are interested in
transforming wave functions $\psi(X)$ to $\psi(\chi)$ the
appropriate choice is to use a function $S(X,\chi)$ of the real
coordinate $\chi^I$ and the complex coordinate $X^I$. It is
determined by requiring that the canonical transformation
\canonicaltransformation\ takes the form
$$
\eta_I={1\over i\pi} {\del S(X,\chi)\over\del \chi^I},\qquad\quad
\bar{X}^I={2\over i\pi}\left({1\over {\rm Im}\tau}\right)^{IJ}{\del
S(X,\chi)\over \del X^J}.
$$
After a little algebra one finds
\eqn\generatingfunction{S(X,\chi)={i\pi \over 2}\left(\chi^I
F_I(X)-{1\over 2}F_0(X)+{1\over 2}\bar F_0(2\chi\!-\!X)\right).}
This leads to the equations
$$
\eta_I={1\over 2}F_I(X)\!+\!{1\over 2}\bar
F_I(2\chi\!-\!X),\qquad\qquad \bar X^I=2\chi^I\!-\!X^I,
$$
which are equivalent to \canonicaltransformation. The relation
between the semi-classical wave function $\psi(X) =e^{g(X)}$ and the corresponding wave function
$\psi(\chi) =e^{f(\chi)}$ in the real polarization is
\eqn\wavefunctions{\psi(X)=\int \!d\chi\, 
e^{{i\over \pi}S(X,\chi)}\psi(\chi),}
where it is understood that the right-hand side is computed in the
saddle point approximation. In terms of the functions $g(X)$ and
$f(\chi)$ this gives the relation
\eqn\canonicalmap{g(X)= f(\chi)+{i\over \pi}S(X,\chi),}
where $\chi^I$ is solved in terms of $X^I$ through
\eqn\condition{{\del f(\chi)\over\del \chi^J}+{i\over \pi}{\del
S(X,\chi)\over\del \chi^J}=0,}
We want to determine the function $f(\chi)$ that corresponds to
the simple wave function $\psi_{0,0}(X)=1$. In that case $g(X)=0$
and hence by differentiating \canonicalmap\ one gets
$$
{\del \over \del X^J}\left(f(\chi(X))+{i\over
\pi}S(X,\chi(X))\right)=0,
$$
where $\chi(X)$ is solved from the condition \condition.  By
inserting the expression \generatingfunction\ on the right-hand side one
finds that this equation obeyed when $X^I=2\chi^I$. Putting this
back in to \canonicalmap\ with $g(X)=0$ finally gives
$$
f(\chi)=\chi^IF_I(\chi)-F_0(\chi)=F_0(\chi),
$$
where we used homogeneity of $F_0(\chi)$. So we conclude that the
canonical transformation maps the wave function $\psi_{0,0}(X)=1$
on to
$$
\psi_{0,0}(\chi)=e^{F_0(\chi)}.
$$
This is precisely the semi-classical approximation to the
topological partition function!

\newsec{Topological Strings and the Exact
Hartle-Hawking Wave-function}

So far we have argued using semi-classical reasoning the form of
the Hartle-Hawking wave function. The aim of this section is to
propose an exact such wave function which agrees in the
semi-classical limit with the wave function discussed before.  We
will argue that the state with no flux can be identified with the
topological string wave function:
$$|\psi_{0,0}\rangle =|\psi_{\rm top}\rangle .$$
That this relation could hold presupposes that
 topological string partition function also corresponds to a wave function
associated to quantizing $H^3(M)$.  This is indeed the interpretation of \Witten\
of the holomorphic anomaly for topological strings \bcov .
  Moreover
this is in agreement with the fact that at least semi-classically
 the topological string partition function
describes the Hartle-Hawking wave function in the real
polarization as discussed in the previous section. Moreover we
argue that the semi-classical result
$$|\psi_{p,q}\rangle =O_{p,q}|\psi_{0,0}\rangle $$
also holds with the expected factor
\eqn\expec{O_{p,q}=\exp\left[ -{i\pi \over 4} W_{p,q}
+c.c.\right]}
interpreted as an operator acting on the topological string Hilbert space.
Important evidence for this proposal comes
from the connection between the black hole entropy and the
topological string partition function discovered in \osv.
To motivate this more clearly let us first briefly review the result of
\osv .

Semi-classically, the entropy of the BPS black hole obtained
by wrapping D3 branes around cycles of $M$ is given by the
area \spherearea\ of the horizon.
It was pointed out in \osv , based
on earlier work \dewit , that the resulting quantum corrected
entropy formula can be concisely expressed as
\eqn\qentropy{ S_{{\rm BH}}(q,p) =  {\cal F}(p, \phi)+ \sum_I q_I
\phi^I, }
where
\eqn\whatsF{ {\cal F}(p, \phi) = F_{\rm top}(X) + \bar{F}_{\rm
top}(\bar X), ~~~X = p + {i \over \pi} \phi.}
and
$$F_{{\rm top}}(X)=\sum_{g=0}^{\infty} F_g(X)$$
is the full topological string partition function.

Moreover the quantum corrected attractor equations also take the
simple form:
\eqn\qattractor{ q_I = - {\partial \over \partial \phi^I}{\cal
F}(p, \phi) .}
At the attractor point, the string perturbation expansion
 is an asymptotic expansion for large black hole
charges. Since \qentropy\ takes the Legendre transformation from
$\phi$ to $q$, it was conjectured in \osv\ that the number of
states $\Omega(p,q)$ of the black hole with finite charges $p^I,
q_I$ is given by  Laplace transformation of the topological string
partition function,
\eqn\exactentropy{ \eqalign{ \Omega(p,q) & = \int d\phi \ e^{- q_I
\phi^I + {\cal F}(p, \phi)} \cr & = \int d\phi \ e^{-q_I\phi^I }
\left|e^{F_{\rm top}\left(p + {i \over \pi} \phi\right)}
\right|^2,}}
More precisely, the conjecture states that $\Omega(p,q)$ given by
\exactentropy\ is the Witten index for the quantum Hilbert space
of the black hole.
This conjecture has been successfully tested\foot{In addition
there are interesting non-perturbative contributions discovered
in these papers which suggest how to define topological
strings beyond perturbation theory.} in examples where
both $\Omega(p,q)$ and $F_{\rm top}(X)$ can be computed
independently and explicitly \refs{\vafaqcd, \aosv,
\atishone }\foot{See also \atishtwo. 
It should be reminded that the expression 
\osv\ is proposed to be 
valid for the perturbative expansion, as 
stressed in \refs{\vafaqcd, \aosv}. Thus, the test of the conjecture 
\osv\ has to be made in the perturbative regime where the topological 
string coupling constant is small. 
}.

We note that the relation \exactentropy\ is an example of the
large $N$ duality. The near horizon limit in the above is
equivalent to the low energy limit for the open string system on
D3 branes wrapping 3-cycles on $M$ and extended in one
time-dimension, and the entropy counts the number of ground
states of this system (with the sign $(-1)^F$). The conjecture
\exactentropy\ states that this can also be evaluated using the
{\it closed} topological string theory on $M$. In principle the
gauge theory on the D3-brane  provides a non-perturbative
holographic definition of topological string theory.

As noted in \osv\ the expression \exactentropy\ for the number of
states can be written in a nice way as a Wigner function.
 Namely, by taking the contour of the $\phi$ integral in \exactentropy\ along
the imaginary axis as $\phi = - i \pi \chi$, one gets
\eqn\wignerfunction{\Omega(p,q)=\int d\chi \ e^{i\pi
q\chi}\bar{\psi}_{{\rm top}}\left(p-\chi\right)
\psi_{\rm top}\left(p+\chi\right),}
where
\eqn\wavefunction{\psi_{\rm top}(\chi)=e^{F_{{\rm top}}(\chi)}}
is the exact topological string partition function.

To relate this result to the normalization of the Hartle-Hawking
wave function, let us  start with the following observation. The
Witten index $\Omega(p,q)$ for the black hole can be evaluated by
analytically continuing the near horizon geometry to  $M\times S^2
\times
 ({\rm Euclidean}~ AdS_2)$, by periodically identifying
the Euclidean time $\tau$ by period $\beta$, and by computing the
partition function with the supersymmetry preserving boundary
condition around the $\beta$-circle. The result should be
independent of $\beta$.

\bigskip
\centerline{\epsfxsize 2.5truein\epsfbox{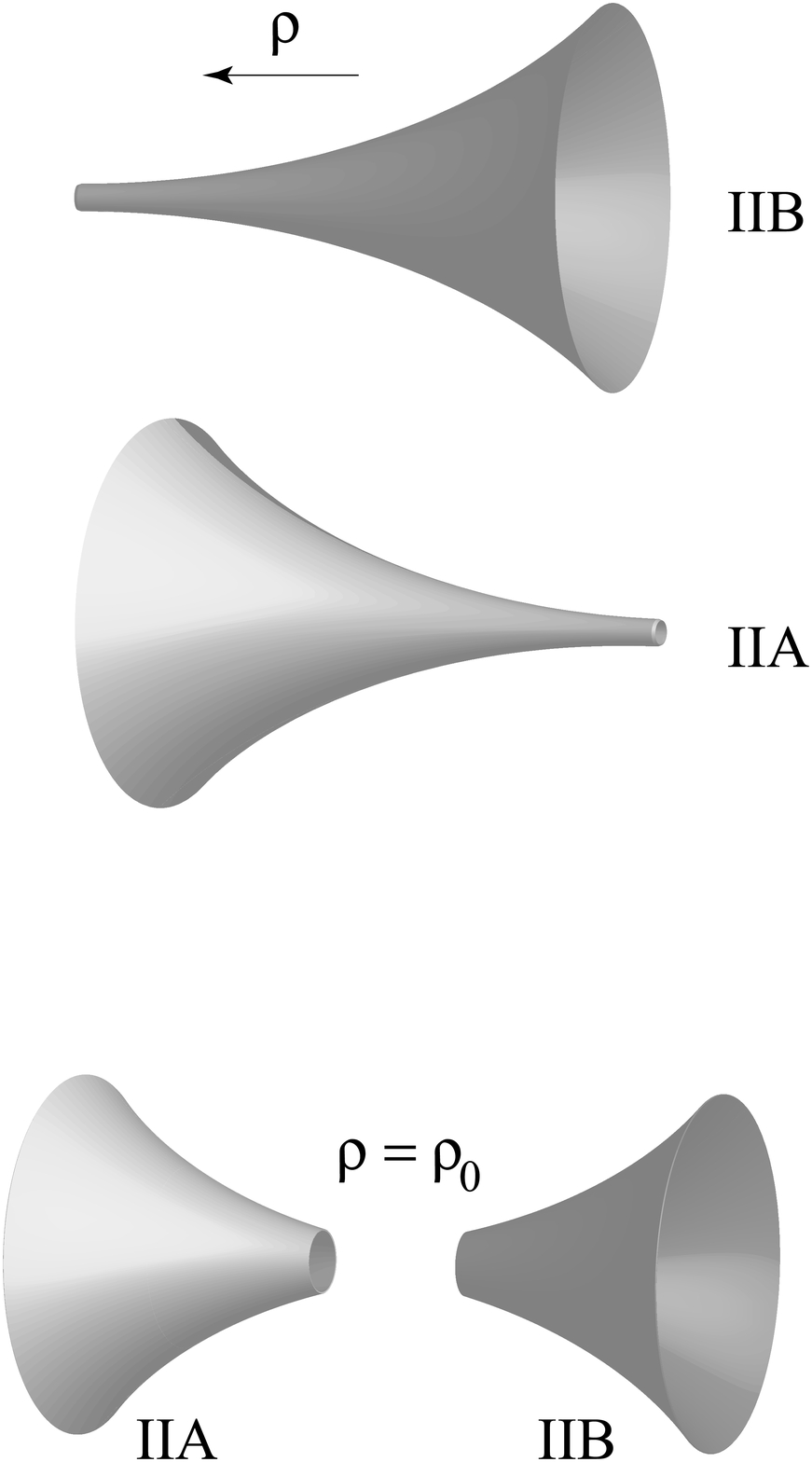}} \leftskip 2pc
\rightskip 2pc \noindent{\ninepoint\sl \baselineskip=2pt {\bf
Fig.2} {{The periodically identified Euclidean $AdS_2$ and
its T-dual are shown in the above. We can combine the
right-hand half of the IIB picture and the left-half of
the IIA picture so that the geometric description is
valid at both sides.}}}
\bigskip

The Euclidean $AdS_2$ with the periodic identification, 
$i.e.$ $H_2/{\bf Z}$, has the topology of a cylinder.  At
one end the radius goes to infinity and at the other end it approaches
zero.  Now, separate the cylinder into
two parts by cutting it across at $\rho=\rho_0$ for some
fixed $\rho_0$. We want to associate a wave function with both sides.
The side at which the radius goes to infinity satisfies the 
quantum BPS equation corresponding to the `upward' attractor flow,
while the other side at which the radius goes to zero is represented 
by a conjugate state corresponding to the  `downward' attractor flow.

One could argue that the geometric description of the solution for radii much
smaller than string length is not trustable, and than should go to the
T-dual description.  In this case we end up with a Euclidean IIA theory
where radius is increasing. On the original IIB side, we have the
near horizon geometry of the D3 branes wrapping 3-cycles in
the Calabi-Yau 3-fold and wrapping the $S^1$. On the IIA side,
we have Euclidean D2 branes wrapping the same 3-cycles in 
the Calabi-Yau and smeared over the dual $S^1$. The mini-superspace
wave-functions on both sides are given by the topological
string partition function since the geometric set-ups in the
Calabi-Yau 3-fold are the same. Thus, viewed in this way,
our Euclidean geometry  is separated into
two parts, the IIB part for $\rho > \rho_0$
and the IIA part for $\rho < \rho_0$,  as shown in Figure 2.  This is similar to the
Euclidean $AdS_2$ periodically identified in the global
time.  The metric is given by \adsmetric\ in the IIB side. 

This naturally suggests an
interpretation of the Witten index ${\rm Tr}(-1)^F e^{-\beta H}$
as an inner product in the Hilbert space for the space $M\times
S^2 \times S^1$. As we mentioned earlier, this is
analogous to the familiar worldsheet duality in the perturbative
open string theory where an annulus diagram
can be represented either as a trace over an open string Hilbert
space or an inner product of boundary states. The difference is
that we are performing this in the target space.   

Let $|\psi_{p,q}\rangle$
denote the state we obtain upon doing the path-integral on the right
in a fixed flux sector.  The above consideration leads to the statement that
$$\Omega(p,q)=\langle \psi_{p,q}|\psi_{p,q}\rangle .$$
To make contact between the discussion in section 5  and the
topological string let us write the expression \exactentropy\ in
invariant form
$$
\Omega(p,q)=\langle\bar{\psi}_{{\rm top}}|e^{i\pi
(q\chi-p\eta)}|\psi_{\rm top}\rangle ,
$$
where $\chi$ and $\eta$ are to be regarded as operators, and the
state $\bar{\psi}$ is defined by the wave function
$\psi^*(-\chi)$. Here, the state on the left is not simply the complex conjugate but it contains a minus sign
due to the time reversal in the attractor flow equation. 
Next we note that  the attractor relations
\canonicaltransformation\ imply that $q_I\chi^I-p^I\eta_I= {\rm
Re}\ W(X)$, and hence
$$
\Omega(p,q)=\langle\bar{\psi}_{{\rm top}}
|e^{{-\pi\over 2}i \left(W(X)-\bar{W}(\bar{X})
\right)}|\psi_{\rm top}\rangle .
$$
  
It follows that if we identify
\eqn\wavfu{|\psi_{p,q}\rangle =e^{-i{\pi\over 4}\left( W(X)-\bar{W}(\bar{X})
\right)}|\psi_{\rm top}\rangle ,}
or in the wave function form as
$$\psi_{p,q}(\phi^I)=e^{-{1\over 2}q_I\phi^I +F_{{\rm top}}
(p^I+{i\over \pi}\phi^I)},$$
then
$$\Omega(p,q)=\langle \psi_{p,q}|\psi_{p,q}\rangle
=\int d\phi^I \ |\psi_{p,q}(\phi^I)|^2$$
exactly as expected.  Moreover the form of the wave function
\wavfu\ is exactly consistent with the semi-classical reasoning
which led to \expec .  The fact that the wave functions for both
the IIA and IIB side would lead to the same state is clear once we
recall that the internal part of the Calabi-Yau and thus the mini-superspace
 is identical for
both cases where a D2 brane IIA instanton is playing the role of
D3 brane of IIB. We find this a highly non-trivial evidence for
our conjecture for the exact Hartle-Hawking wave function. 

 Note
that even though for many purposes the semi-classical wave function
we presented earlier may suffice, the very fact that in a
consistent quantum theory of gravity the exact quantum corrected
Hartle-Hawking wave function could be made sense out of is a
rather significant statement.

\bigskip

\noindent
{\it Holomorphic versus Real Polarization}

We end this section with a brief comment on the issue of normal
ordering in the K\"ahler quantization in terms of the $(X,\bar{X})$
coordinates and its connection to the holomorphic anomaly
equations.  We also make contact with the work in \EV\ which
reformulates the black hole entropy proposal of \osv\ in a background independent way.

 To give a precise definition of the K\"ahler quantization
procedure discussed in section 5 one has to resolve the operator ordering
ambiguities that arise due to the non-linearity of the canonical
transformation \canonicaltransformation\ from $(\chi,\eta)$ to
$(X,\bar{X})$. In the discussion above we have ignored this issue,
so part of our arguments are semi-classical. In particular, the
relation between the wave function $\psi_{\rm top}(X)$ and the exact
topological string partition function $\psi_{\rm top}(\chi)$
\eqn\topwavefunctions{\psi_{\rm top}(X)=\int \!d\chi\, e^{{i\over
\pi}S(X,\chi)}\psi_{\rm top}(\chi)}
is a semi-classical formula. Nevertheless, the topological string
is well defined and its partition function exists to all orders in
perturbation theory. Hence, in principle it describes the loop
corrections to the Hartle-Hawking wave function to all orders in
perturbation theory.

The partition function of topological string theory can be
computed perturbatively around a given background by writing
$X^I=Z^I+x^I$, and treating the perturbation $x^I$ as coupling
constants on the worldsheet\foot{The real polarization
corresponds to setting the background $Z^I$ at infinity where
the magnetic $B$ cycles are infinitely bigger than the electric $A$ cycles.}.
It is well known that the resulting
partition function $\psi_{\rm top}(x;Z,\bar{Z})$ depends
holomorphically on $x^I$ but has a non-holomorphic dependence on
the background due to holomorphic anomaly. One of the original
motivations for interpreting $\psi_{\rm top}$ as a wave function is
that it naturally explains this background dependence \Witten.
Namely, it arises because the topological string partition
function is the wave function of the state $|\psi_{\rm top}\rangle$ in
a complex polarization that is determined by the background. In
fact, the coordinates $x^I$ used by the perturbative topological
string are a linearization of the `curved' $X^I$ coordinates. To
be precise, the relation with $(\chi,\eta)$ is
\eqn\linearized{\chi^I={\rm Re}\left(Z^I+x^I\right)\qquad
\eta_I={\rm Re}\left(F_I(Z)+\tau_{IJ}x^J\right),}
where $\tau_{IJ}=\del_I\del_J F_0(Z)$ is determined by the
background. These are just  the attractor equations
\canonicaltransformation\ linearized around $X^I=Z^I$. In this way
the topological string avoids the normal ordering problems but at
the cost of a background dependence. The partition function
$\psi_{\rm top}(x;Z,\bar Z)$ is related to the background independent
wave function $\psi_{\rm top}(\chi)$ via the Bargmann transform
\refs{\DVV,\EV}
\eqn\transform{ \psi_{\rm top}(x;Z,\bar{Z})=|{\rm det}{\rm
Im}\tau|^{1\over 2} \int d\chi \ e^{iS(x,\chi;Z,\bar
Z)}\psi_{\rm top}(\chi),}
with
$$
S(x,\chi;Z,\bar Z)={\pi\over
4}\bar{\tau}_{IJ}(\bar{Z})\chi^I\chi^J\!+\pi
\chi_I(x^I\!+\!Z^I)\!+\!
 {\pi\over 4} (x^I \!+\! Z^I)(x_I \!+\! Z_I),
$$
where indices are lowered with ${\rm Im}\tau_{IJ}(Z)$. This
expression is a linearization of the generating function
$S(X,\chi)$ defined in \generatingfunction. The topological
partition function $\psi(x;Z,\bar{Z})$ thus gives in a certain
sense a linearized description of the Hartle-Hawking wave function
$\psi_{\rm top}(X)$. We like to stress, however, that it is not a
`linear approximation': it is an exact wave function that carries
the complete information about the state $|\psi_{\rm top}\rangle$, and
hence about the Hartle-Hawking wave function.

\newsec{More General Compactifications}

It is clear that the above philosophy, $i.e.$ finding a wave function
for string compactifications should be a generally valid
principle when we compactify all dimensions.  So in particular
we should aim to write a wave function for the string vacuum
upon compactification of type II superstrings on $X^9$ or for M-theory
on $X^{10}$.  There are many interesting cases to consider, for example
when $X^9=X^6\times S^3$ where $X^6$ is a Calabi-Yau and $S^3$ has some
fluxes through it.  One could also consider the case where $X^9=S^5\times S^4$;
this may be relevant in the context of the usual $S^5\times AdS_5$ compactifications
of type IIB.  We leave a study of such wave functions to future work. Instead
here we consider a particular case motivated by the example of 5 dimensional
black holes obtained upon compactifications of M-theory on a Calabi-Yau threefold $X^6$.

Consider $M$-theory compactified on a spatial geometry $X^6\times S^3\times S^1$,
where we turn on the 7-form field strength flux $*G$ as
$$*G=F_4\wedge \omega ,$$
where $\omega$ is the unit volume form on $S^3$ and $F_4$ is an integral four-form
on $X^6$, which is dual to an integral two cycle class $Q\in H_2(X^6,{\bf Z})$.
The setup is very similar to the main setup of this paper where
we can view this as a circle compactification of the Euclidean 5d black hole
geometry $X^6\times S^3\times AdS_2$.  This is holographically dual
to computation of the black holes corresponding to $M2$ branes wrapping
2-cycles of $X^6$.  Just as in our main example, we can write the mini-superspace
wave function. In this case the mini-superspace will involve K\"ahler structure
of $X^6$ which we denote by $k$ and flux through $S^3$ which we denote by $F_4$.
It will not depend on the complex moduli of Calabi-Yau, nor on the radius of $S^1$.
A reasoning very similar to our discussion for 4d black holes leads us to
$$\Psi(k,F_4)={\rm exp}
\left(-{\pi\over 12}\int_{X^6}  k\wedge k\wedge k +{\pi\over 2} \int_{X^6} k\wedge F_4\right) .$$
This was motivated as follows:  As discussed before, by holography
this wave function should have the property that
$$\Omega(Q)=\int dk \left|\Psi(k,F_4)\right|^2 ,$$
and
by the recent observation in
\topm\ the leading saddle point computation gives
$$\Omega(Q)={\rm exp}\left({\pi \sqrt{2} \over 3}\int_{X^6} F_4^{{3\over 2}}\right) ,$$
which indeed agrees with the  semi-classical expectation for the entropy.  Moreover
 the wave function is peaked at
$${1\over 2}k\wedge k =F_4 ,$$
which is the attractor equation for 5d black hole, fixing the K\"ahler moduli
in terms of the flux.

 The above
wave function is expected to be true to all orders in the $1/N$ expansion, but
not beyond that, just as in the 4d black hole case discussed in this paper.  It also
suggests that the entropy as a function of charge $Q$ is an Airy function, to all orders
in the $1/Q$ expansion.  It would be interesting to check this.

\newsec{Applications}

We will end this paper with discussion on some directions
for future research. 

\subsec{Connection with Cosmology:  Attaching to Lorentzian Signature Spacetime?}

The purpose of the original Hartle-Hawking construction of the
wave-function is to use it to set up an initial condition for a
universe with a positive cosmological \hh. In the previous sections, we
argued that the topological string partition function gives an
analogous wave-function when the cosmological constant is
negative.  Since the Hilbert space is common between Lorentzian
and Euclidean signature, we can view our state $|\psi\rangle$ as a state
in the Lorentzian signature mini-superspace Hilbert space.  It would
then be natural to evolve this in time.   This is, however, not
as trivial as it my appear.  The reason for this is that
the Euclidean signature metric on $H_2/{\bf Z}$ is given by
\eqn\euclideanmetric{ ds_E^2 = d\rho^2 + e^{2\rho} d\tau^2 ,}
where $\tau$ is periodically identified as $\tau \sim \tau+\beta$.
In the previous section, we regarded $\rho$ as the Euclidean time of
this geometry and constructed the wave-function $\Psi$. If we
 cut the Euclidean geometry at some value of $\rho$, and attempt
to glue a Minkowski solution by analytic continuation
 $\rho=it$ the metric would not be real.  Nevertheless it is clear
that we have a well defined state in the Hilbert space of the Minkowski
theory in the sense of mini-superspace, and thus it should give
rise to some evolution in Lorentzian signature.  This issue is currently
under investigation.

\subsec{Statistical Interpretation of Topological String Wave-Function}

We have found a wave function for the mini-superspace in the context
of flux compactification of type IIB on $S^1\times S^2\times M$, where
the mini-superspace consists of the choice of flux $(p,q)$ as well
as the extended complex moduli of $M$ (which includes the overall radius of $S^2$).
We can use the wave function $\Psi_{p,q}(x)$ to measure the probability
 distribution
for various fluxes and moduli.
Note that we {\it can} compare the different fluxes here, as explained in
section 3, because finite action instantons do interpolate between
different flux sectors.

{}From this result and the fact that the norm of the wave function
is the exponential of the entropy, we are immediately led to the
conclusion that the large fluxes, leading to more entropy, are
most likely.  In other words the flat space, corresponding to
$p,q\rightarrow \infty$ is the most likely situation. Thus in this
class of supersymmetric theories we would predict that the most
likely situation is for the universe to be flat. This is very much
in the spirit of Hartle-Hawking \hh .  Note that this prediction
is very different from the prediction one would have made based on
counting the number of classical solutions.  That would have led
to equal probability for each flux $(p,q)$, as there is exactly
one solution for each choice of flux.

We could also have asked the following question:  Suppose
we {\it know}
that flux is a given value $(p,q)$, $i.e.$
that has been already measured.  Then
what is the probability of finding various $X^I$?  Then we can use the wave function
in that sector and the conclusion would be that it is peaked at the attractor
value, as already discussed.  We can ask how strongly peaked is it?  From the
leading part of the wave function (including the measure) which is
$$\exp|X-X_{p,q}|^2,$$
we can conclude that the standard deviation away from the
attractor value is of order $1$, $i.e.$,
$$\Delta X^I \sim 1,$$
This leads to the statement that the deviation of the Calabi-Yau
moduli away from the attractor value is given by
$${|\Delta X|\over |X|}\sim {1\over Q},$$
where $Q$ denotes the overall scale of the charge.  For small
charge $Q$, the moduli of Calabi-Yau is not very peaked near the
attractor value; however for large $Q$ it becomes increasingly
more peaked.  This is consistent with the fact that for large $Q$
the radius of $S^2$ scales as $Q$ and so we end up in non-compact
space in that limit, which should freeze the moduli of Calabi-Yau.


It is quite gratifying to see that we can compare various string compactification
data if we compactify {\it all} spatial moduli.  It would be very
useful to find more such examples within string theory and move toward
a more realistic quantum cosmology within string theory.

\bigskip
\medskip
\centerline{\bf Acknowledgments} We would like to thank F.~Denef,
R.~Dijkgraaf, G. Gibbons, R.~Gopakumar, S.~Gukov, G.~Mandal,
S.~Minwalla, L.~Motl, A.~Neitzke, J.~Preskill, A.~Strominger, and L.~Susskind
for useful discussions.

The research of H.O. was supported in part
by DOE grant DE-FG03-92-ER40701.
The research of C.V. was supported
in part by NSF grants PHY-0244821 and DMS-0244464.
\listrefs
\end